\documentclass[11pt,a4paper]{article}
\usepackage{amsmath}

\usepackage[psamsfonts]{amssymb}
\usepackage{amsmath}
\usepackage{epsfig}

\author{H. Mohseni Sadjadi\footnote{mohsenisad@ut.ac.ir}
\\ {\small Department of Physics, University of Tehran,}
\\ {\small P. O. B. 14395-547, Tehran 14399-55961, Iran}}
\title{On coincidence problem in ELKO dark energy model}

\begin{document}
\maketitle
\begin{abstract}
We study the critical points of a universe dominated by ELKO
spinor field dark energy and a barotropic matter without
considering a specific potential or interaction. The coincidence
problem and attractor solutions are discussed at late time, and it
is shown that the coincidence problem can not be solved in this
model. \vspace{0.5cm}\newline
\end{abstract}

\section{Introduction}

To describe the present accelerated expansion of the universe
\cite{acc}, many models have been considered. In dark energy
models, almost 70\% of our universe is assumed to be filled with a
smooth unknown matter with negative pressure known as dark energy.
The first candidate for dark energy was the cosmological constant
which may arise from quantum vacuum energy density \cite{wein}.
Other candidates proposed for dark energy were exotic dynamical
scalar fields such as quintessence or phantom, but, spinor dark
energy model has also attracted some attentions recently
\cite{spin}.

In \cite{spin1}, a class of non standard spinors, constructed in
momentum space from the eigenspinors of the charge conjugation
operator, known as ELKO spinor(Eigenspinoren des
LadungsKonjugationsOperators), satisfying $(CPT)^2=-1$, was
introduced and proposed as dark matter candidate . Cosmological
and gravitational consequences of this model were studied in
\cite{spin2}.

Although, at the beginning, ELKO spinor was contemplated as dark
matter, but subsequently, it was considered as a potential
candidate for inflation \cite{inf}, or present acceleration of the
universe \cite{darken}. This last point is the subject of the
present paper.

Viable models must be consistent with astrophysical data. These
data indicate that, despite the expansion of the universe, the
ratio of matter to dark energy density is of order
$r:={\rho_m\over \rho_d}\simeq \mathcal{O}(1)$. This problem, i.e.
why dark energy density ($\rho_d$) is of the same order of the
matter density  ($\rho_m$), is known as the coincidence problem
\cite{coin}.

In \cite{spin3}, it was found that, for some special potentials
and also for some special interactions between (dark) matter and
dark energy, the coincidence problem cannot be solved in ELKO dark
energy model.

In this paper, we consider the late time evolution of the universe, which is assumed
to be almost composed of a barotropic matter and ELKO dark energy.
By considering a general dark
energy potential and also a general interaction between
matter and dark energy, it is shown that, in principle, the
coincidence problem cannot be alleviated in this model.

We use the units $\hbar=c=1$.

\section{Attractor solutions in ELKO cosmology}
ELKO model in a curved space is characterized by the action
\begin{equation}
S=\int\left[{1\over
2}{\stackrel{\neg}{\psi}}\overleftarrow{\nabla_{\nu}}\nabla^\nu
\psi+ V({\stackrel{\neg}{\psi}}\psi)\right]\sqrt{-g}d^4x,
\end{equation}
where $g$ is determinant of the metric tensor, $V$ is the
potential, and ${\stackrel{\neg}{\psi}}$ is dual spinor.
$\nabla_\nu$ are covariant derivative components acting on spinors
as
\begin{equation}
\nabla_\nu\psi=(\partial_\nu-\Gamma_\nu)\psi,
\end{equation}
where $\Gamma_\nu={i\over 4}\omega_{\nu ab}\sigma^{ab}$.
$\omega_{\nu ab}$ are spin connection components, and in terms of
Dirac matrices we have  $\sigma^{ab}={i\over
2}\left[\gamma^a,\gamma^b\right]$. The energy momentum tensor is
obtained as (the derivation of energy momentum tensor of ELKO
spinors can be found, in details, in \cite{spin2})
\begin{eqnarray}
&&T^{\mu \nu}{\stackrel{\neg}{\psi}}\overleftarrow{\nabla^{(\mu}}\nabla^{\nu)}\psi-g^{\mu
\nu}\left[{1\over
2}{\stackrel{\neg}{\psi}}\overleftarrow{\nabla_{\alpha}}\nabla^{\alpha}\psi-V({\stackrel{\neg}{\psi}}\psi)\right]
\nonumber \\
&&-{i\over
4}\nabla_{\beta}\left[{\stackrel{\neg}{\psi}}\overleftarrow\nabla^{(\mu}\sigma^{\nu)\beta}\psi+
{\stackrel{\neg}{\psi}}\sigma^{\beta(\mu }\nabla^{\nu
)}\psi\right].
\end{eqnarray}
In the following we consider the spatially flat FRW space-time
\begin{equation}\label{1}
ds^2=dt^2-a^2(t)(dx^2+dy^2+dz^2),
\end{equation}
where $a(t)$ is the scale factor. In this background by writing
ELKO spinor  as $\psi= \phi \lambda$, where $\lambda$ is a
constant spinor satisfying ${\stackrel{\neg}{\lambda}}\lambda=1$
\cite{spin2}, the energy momentum tensor may be derived as
\begin{eqnarray}\label{2}
T^0_0&=&\rho_d={1\over 2}\dot{\phi}^2+V+{3\over 8}H^2\phi^2\nonumber \\
T^i_j&=&-\delta^i_j P_d=\delta^i_j\left({3\over
8}H^2\phi^2+V-{1\over 2}\dot{\phi}^2+{1\over
4}\dot{H}\phi^2+{1\over 2}\phi \dot{\phi}H\right).
\end{eqnarray}
$\rho_d$, and $P_d$ are the energy density and pressure of dark
energy respectively. In the absence of interaction, the continuity
equation for the dark sector
\begin{equation}\label{3}
\dot{\rho}_d+3H(P_d+\rho_d)=0,
\end{equation}
implies
\begin{equation}\label{4}
\ddot{\phi}+3H\dot{\phi}+{dV\over d\phi}-{3\over 4}H^2\phi=0.
\end{equation}

Now let us consider a FRW Universe dominated by ELKO spinor dark
energy and a barotropic matter $\rho_m$ whose the pressure is
$P_m=w_m\rho_m$. The equation of state parameter of the matter,
$w_m$,  is assumed to be non negative  $w_m\geq 0$, e.g. for cold
dark matter we have $w_m=0$. The continuity equations for dark
energy and matter component in the presence of the interaction
source $C$ become
\begin{eqnarray}\label{5}
&&\dot{\rho_d}+3H(P_d+\rho_d)=-C\nonumber \\
&&\dot{\rho_m}+3H\gamma \rho_m=C,
\end{eqnarray}
where $\gamma:=w_m+1$. The Friedmann equation is given by
\begin{equation}\label{6}
H^2={1\over 3M_p^2}(\rho_d+\rho_m),
\end{equation}
which can be rewritten as
\begin{equation}\label{7}
(1-{1\over 8M_p^2}\phi^2)H^2={1\over 3M_p^2}(\rho_m+{1\over
2}\dot{\phi}^2+V).
\end{equation}
$M_p$ is the reduced Planck mass. So the effective gravitational
coupling constant is modified in this theory. For $(\rho_m+{1\over
2}\dot{\phi}^2+V)>0$, we must have $|\phi|<2\sqrt{2}M_p$.

Raychaudhuri equation reads:
\begin{equation}\label{8}
(1-{1\over 8M_p^2}\phi^2)\dot{H}=-{1\over
2M_p^2}(\gamma\rho_m+\dot{\phi}^2-{1\over2}H\phi\dot{\phi}).
\end{equation}
The scalar field, $\phi$, satisfies the classical equation of
motion:
\begin{equation}\label{9}
\dot{\phi}\left(\ddot{\phi}+3H\dot{\phi}+V_{,\phi}-{3\over
4}H^2\phi\right)=-C.
\end{equation}
The equation of state parameter of the Universe defined by
$w={{P_d+P_m}\over {\rho_d+\rho_m}}$ is given by $w=-1+{2\over
3}\omega$, where $\omega=-{\dot{H}\over H^2}$.

To study the cosmological dynamics of this model, we define
dimensionless variables \cite{dimensionless}
\begin{equation}\label{10}
x={\dot{\phi}\over \sqrt{6}M_pH},\,\,\,y={\sqrt{V}\over
\sqrt{3}M_pH},\,\,\,z={\sqrt{\rho_m}\over \sqrt{3}M_pH},\,\,
u={\phi \over M_p\sqrt{8}}.
\end{equation}
Hence
\begin{equation}\label{100}
(1-u^2)\omega=3x^2+1.5 \gamma z^2-\sqrt{3}xu.
\end{equation}

By the assumption that the potential is only a function of $u$,
and by defining
\begin{equation}\label{11}
f(u)={M_pV_{,\phi}\over V},
\end{equation}
where $V_{,\phi}={dV\over d\phi}$, we find out the autonomous
system of differential equations
\begin{eqnarray}\label{12}
x'&=&(\omega-3)x+{\sqrt{3}\over 2}u-\sqrt{3\over 2}y^2f-C_1\nonumber \\
y'&=&(\sqrt{3\over 2}xf+\omega) y \nonumber \\
z'&=&(\omega-{3\over 2}\gamma)z+C_2 \nonumber \\
u'&=&{\sqrt{3}\over 2}x,
\end{eqnarray}
where, prime denotes derivatives with respect to the e-folding
time  $\mathcal{N}=\ln{a}$, and
\begin{eqnarray}\label{13}
C_1&=&{C\over \sqrt{6}M_pH^2\dot{\phi}}\nonumber \\
C_2&=&{C\over 2\sqrt{3}M_pH^2\sqrt{\rho_m}}={x\over z}C_1.
\end{eqnarray}
$C$ is taken to be a function of $x,z,u$. For the sake of
generality, we do not restrict ourselves to a specific
interaction.

Note that $x,y,z,u$ are not independent and are constraint to the
Friedmann equation :
\begin{equation}\label{14}
x^2+y^2+z^2+u^2=1.
\end{equation}

Most generally, critical points of  the autonomous system
(\ref{12}) denoted with $\{\bar{x},\bar{y},\bar{z},\bar{u}\}$ can
be arranged as follows:

I: $\{\bar{x}=0,\bar{y}=0,\bar{z}=0,\bar{u}=0\}$ which is in
contradiction with (\ref{14}), and then is ruled out.

II:$\{\bar{x}=0,\bar{y}=0,\bar{z}\neq 0,\bar{u}=0\}$. From
(\ref{14}) we have $\bar{z}^2=1$. (\ref{12}) implies $\bar
{C_2}=0$ and $\bar{C_1}=0$, where bar denotes the value at the
critical point.

III:$\{\bar{x}=0,\bar{y}=0,\bar{z}=0,\bar{u}\neq 0\}$. From
(\ref{14}) we have $\bar{u}^2=1$, (\ref{12}) gives $\bar {C_2}=0$
and $\bar{C_1}=\pm {\sqrt{3}\over 2}$.

IV:$\{\bar{x}=0,\bar{y}=0,\bar{z}\neq 0,\bar{u}\neq 0\}$. In this
case, $\bar{u}^2+\bar{z}^2=1$, and $\bar{C_1}={\sqrt{3}\over
2}\bar{u}$.  (\ref{100}) gives $\bar{\omega}= {3\over 2}\gamma$.
We have also $\bar{C_2}=0$.

It is obvious that in the absence of interaction critical points
III and IV do not exist.

V:$\{\bar{x}=0,\bar{y}\neq 0,\bar{z}=0,\bar{u}=0\}$,
$\bar{y}^2=1$. From (\ref{12}) we have $\bar{C_1}=-\sqrt{3\over
2}\bar{f}$, $\bar{C_2}=0$, and (\ref{100}) gives $\bar{\omega}=0$.

VI:$\{\bar{x}=0,\bar{y}\neq 0,\bar{z}\neq 0,\bar{u}=0\}$,
$\bar{z}^2+\bar{y}^2=1$. From (\ref{12}) we obtain
$\bar{\omega}=0$ which using $\bar{\omega}=1.5 \gamma \bar{z}^2$,
results in $\gamma=0$. In this case, $\bar{C_1}=-\sqrt{3\over
2}\bar{f}\bar{y}^2$, and $\bar{C_2}=0$.

In the absence of interaction,  critical points V and VI exist
only for potential satisfying $\bar{f}(u)=f(\bar{u})=0$. E.g. for
the potential $V\propto exp(\lambda \phi)$, we have
$\bar{f}(u)=\lambda M_P$, and these critical points do not exist
when $C=0$.

VII: $\{\bar{x}=0,\bar{y}\neq 0,\bar{z}=0,\bar{u}\neq 0\}$,
$\bar{u}^2+\bar{y}^2=1$. From (\ref{12}) we obtain $\bar{C_2}=0$,
$\bar{\omega}=0$ and $\bar{C_1}={\sqrt{3}\over
2}\bar{u}-\sqrt{3\over 2}{\bar{y}}^2\bar{f}$.

VIII: $\{\bar{x}=0,\bar{y}\neq 0,\bar{z}\neq 0,\bar{u}\neq 0\}$.
$\bar{\omega}\bar{y}=0$ gives $\bar{\omega}=0$. We obtain also
$\bar{C_2}={3\over 2}\gamma \bar{z}$, and
$\bar{C_1}={\sqrt{3}\over 2}\bar{u}-\sqrt{3\over
2}{\bar{y}}^2\bar{f}$. Using $C_2={x\over z}C_1$, we obtain
$\bar{C_2}=0$ which results in $\gamma=0$. This could be obtained
in another way : as $\bar{\omega}={3\gamma\over 2}{\bar{z}^2\over
{1-\bar{u}^2}}$, $\bar{\omega}=0$ implies $\gamma=0$. In the
absence of interaction,  critical points VII and VIII exist only
for potential satisfying $\bar{u}=\sqrt{2}{\bar{y}}^2\bar{f}$.

All the critical points are characterized by $\bar{x}=0$, so as it
can be seen from eqs.(\ref{13}) and (\ref{9}), $C_1$ may be
singular or not generally well defined. In theses cases the attractor solutions do
not exist. Instead, if the order of magnitude
of numerator of $C_1$ in (\ref{13}) (i.e. $C$) is less or equal
than that of its denominator at a critical point, $C_1$ is still
finite and well defined. To elucidate this subject, let us
consider interactions $C^{I}=\sigma H\rho_m$ and $C^{II}=\varsigma
H(\rho_m+\rho_d)$, where $\sigma$ and $\varsigma$ are two
constants.  In terms of variables defined in (\ref{10}), $C_1$ can
be rewritten as $C_1^{I}={\sigma\over 2}{z^2\over x}$, which is
not well defined, and $C_1^{II}={\varsigma\over 2x}$ which is
singular at $\bar{x}=0$. Instead, for the interaction
$C^{III}=\alpha{\rho_m\over M_p}\dot{\phi}$, we have
$C_1^{III}=\sqrt{3\over 2}\alpha z^2$ which is well defined. If
one adopts the interaction $C^{III}$, then,  critical points $IV$,
$V$, $VI$, $VII$, $VIII$ are acceptable provided that,
$\sqrt{2}\alpha \bar{z}^2=\bar{u}$, $\bar{f}=0$, $-\bar{f}=\alpha
\bar{z}^2$, $\alpha
\bar{z}^2={1\over{\sqrt{2}}}\bar{u}-\bar{y}^2\bar{f}$, and
$\alpha\bar{z}^2={1\over{\sqrt{2}}}\bar{u}-\bar{y}^2\bar{f}$, hold
respectively.

To study the stability of the system around the critical points,
$II-VIII$ (if exist), we consider small perturbation around these
points, $\{\bar{x},\bar{y},\bar{z},\bar{u}\}\to \{\bar{x}+\delta
x,\bar{y}+\delta y,\bar{z}+\delta z,\bar{u}+\delta u\}$. If the
real part of all the eigenvalues of $\mathcal{M}$ defined by
\begin{equation}\label{15}
{d\over d\mathcal{N}}{ \left( \begin{array}{ccc}
\delta x \\
\delta y \\
\delta u
\end{array} \right)}=\mathcal{M} \left( \begin{array}{ccc}
\delta x \\
\delta y \\
\delta u
\end{array} \right)
\end{equation}
are negative at a critical point, the system has stable attractor
solution. In our model
\begin{equation}\label{16}
\mathcal{M}= \left( \begin{array}{ccc}
\bar{\omega}-3-\bar{C}_{1,x} &
-\sqrt{6}\bar{y}\bar{f}-{\bar{y}\bar{C}_{1,z}\over \bar{z}} &
{\sqrt{3}\over 2}- \sqrt{3\over 2}{\bar{y}}^2 \bar{df\over du}
-\bar{C}_{1,u}-{\bar{u}\bar{C}_{1,z}\over \bar{z}} \\
\sqrt{3\over 2}\bar{y}\bar{f}-{\bar{y}\over
{1-\bar{u}^2}}\sqrt{3}\bar{u} & \bar{\omega}
-{3\gamma \bar{y}^2\over {1-\bar{u}^2}} & {\bar{y}\over {1-\bar{u}^2}}(2\bar{\omega} \bar{u}-3\gamma \bar{u}) \\
{\sqrt{3}\over 2}& 0 & 0
\end{array} \right).
\end{equation}

As we are interested to study the coincidence problem, among the
situations I-VIII, we need only consider cases where $r$ is of
order $\mathcal{O}(1)$ or more precisely:
\begin{equation}\label{17}
r:={\rho_m\over \rho_d}={z^2\over {1-z^2}}\simeq {3\over 7}.
\end{equation}
In the cases III, V, and VII, we have $\bar{r}=0$, and in the case
II, $\bar{r}\to \infty$. Therefore among all the possible critical
points $II-VIII$, only IV, VI, and VIII may be consistent with
 $\bar{r}\sim \mathcal{O}(1)$.

In VI, and VIII, we have $\gamma=0$ which implies $w_m=-1$, in
contradiction with our assumption that the universe is dominated
by ELKO dark energy and a matter with non-negative pressure. So,
finally, we are left only with the case IV. In this case
\begin{equation}\label{18}
\mathcal{M}_{IV}= \left( \begin{array}{ccc} {3\over
2}(\gamma-2)-\bar{C}_{1,x} & 0 & {\sqrt{3}\over 2}
-\bar{C}_{1,u}-{\bar{u}\bar{C}_{1,z}\over \bar{z}} \\
0& {3\over 2}\gamma
 & 0 \\
{\sqrt{3}\over 2}& 0 & 0
\end{array} \right).
\end{equation}
One of the eigenvalues of $\mathcal{M}_{IV}$ is $\lambda={3\gamma
\over 2}$ which is positive, so even in this situation the system
is not stable and there is no scaling attractor. Besides for an
accelerated expanding universe we must have $w<-{1\over 3}$,
implying $\omega<1$, but in IV, we have $\bar{\omega}={3\over
2}\gamma>1$ which, in contradiction with the nowadays accelerated
expansion of the Universe, describes a decelerated expanding
Universe.

\section{Summary}
After the introduction of ELKO spinor as a potential candidate of
dark matter in \cite{spin1},  some attempts has been done to
consider this kind of spinors as inflaton \cite{inf}, and dark
energy \cite{darken}.

In this paper we examined this theory in the context of dark
energy model, and using Friedmann and Raychaudhury equations we
obtained an autonomous dynamical system describing the behavior of
a spatially flat FRW Universe dominated by ELKO non standard
spinor dark energy, interacting with a barotropic matter, at late
time. We did not restrict the problem to special potentials or
interactions. The critical points and attractor solutions of the
problem were studied. The coincidence problem was discussed in
this framework and it was found that there is no stable solution
which can alleviate the coincidence problem.

\end{document}